\documentclass{bmvc2k}
\usepackage{booktabs}
\usepackage{amsmath,amssymb,amsfonts}
\usepackage[utf8]{inputenc}
\usepackage[T1]{fontenc}
\usepackage{float}
\usepackage{graphicx}
\usepackage{xcolor}

\title{CLIP Based Region-Aware Feature Fusion for Automated BBPS Scoring in Colonoscopy Images}

\makeatletter

\renewcommand{\maketitle}{
  \begin{center}
    {\Large\bfseries \@title\par}
    \vspace{1.5em}
    
{\large
  Yujia Fu\textsuperscript{1} \ 
  Zhiyu Dong\textsuperscript{2,*} \ 
  Tianwen Qian\textsuperscript{3} \ 
  Chenye Zheng\textsuperscript{1} \ 
  Danian Ji\textsuperscript{2,†} \ 
  Linhai Zhuo\textsuperscript{1,†}\par
}

\vspace{1em}

{\footnotesize
  \textsuperscript{1}Fuzhou University \quad 
  \textsuperscript{2}Fudan University Huadong Hospital \quad
  \textsuperscript{3}East China Normal University\par
  \vspace{0.2em}
  \textsuperscript{*}Co-first author \quad \textsuperscript{†}Corresponding author\par
}

\vspace{1.5em}
  \end{center}
}

\makeatother

\addauthor{}{}{}
\addinstitution{}{}
\runninghead{Fu et al.}{Region-Aware Feature Fusion for Automated BBPS Scoring}

\begin{document}

\maketitle

\begin{abstract}
Accurate assessment of bowel cleanliness is essential for effective colonoscopy procedures. The Boston Bowel Preparation Scale (BBPS) offers a standardized scoring system but suffers from subjectivity and inter-observer variability when performed manually. In this paper, to support robust training and evaluation, we construct a high-quality colonoscopy dataset comprising 2,240 images from 517 subjects, annotated with expert-agreed BBPS scores. We propose a novel automated BBPS scoring framework that leverages the CLIP model with adapter-based transfer learning and a dedicated fecal-feature extraction branch. Our method fuses global visual features with stool-related textual priors to improve the accuracy of bowel cleanliness evaluation without requiring explicit segmentation. Extensive experiments on both our dataset and the public NERTHU dataset demonstrate the superiority of our approach over existing baselines, highlighting its potential for clinical deployment in computer-aided colonoscopy analysis.
\end{abstract}


\section{Introduction}
\label{sec:intro}

Colonoscopy is widely considered the gold standard for diagnosing and managing gastrointestinal diseases. Its diagnostic accuracy and therapeutic effectiveness are critically dependent on the quality of bowel preparation, which directly influences lesion visibility and procedural success.

\begin{figure}[ht]
  \centering
  \includegraphics[width=0.72\linewidth]{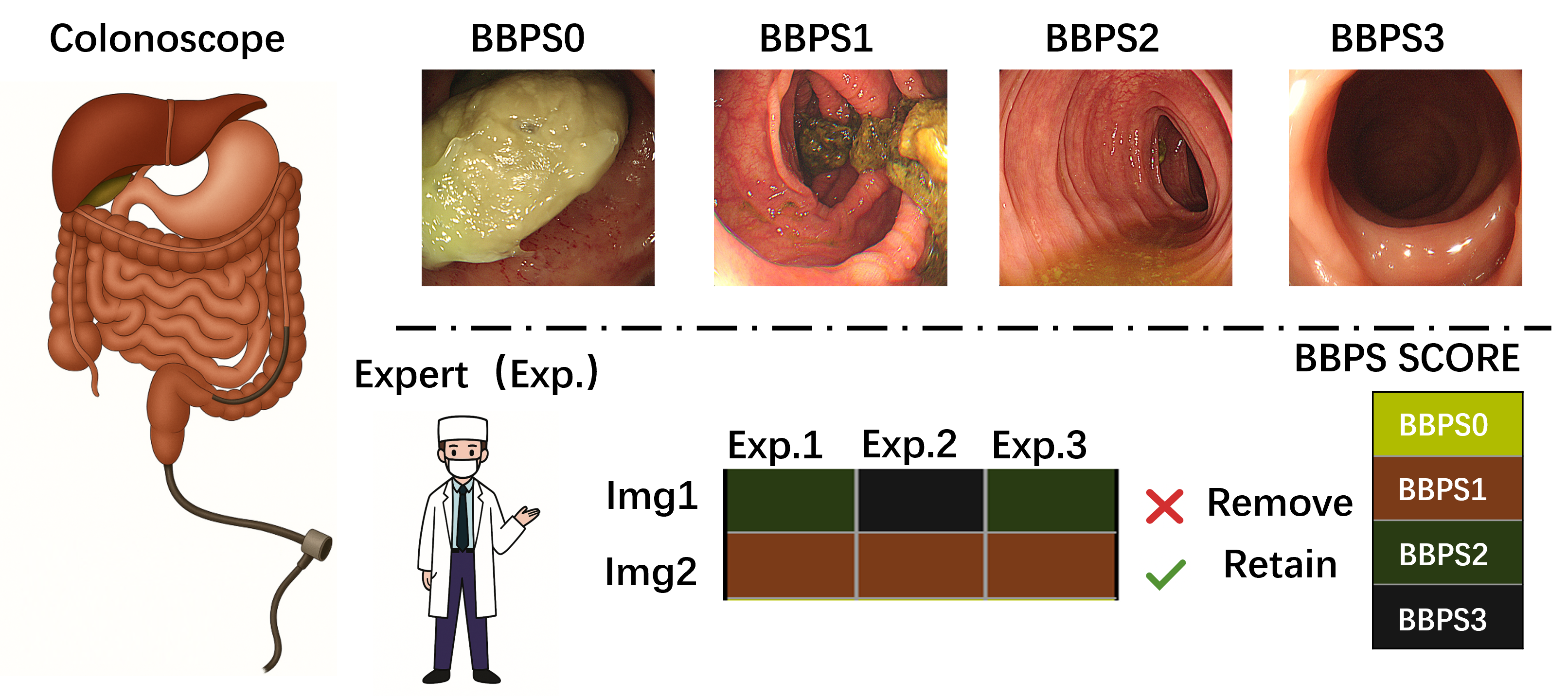}
  \caption{Illustration of BBPS scoring system and our annotation process. 
  Top: Examples of BBPS scores 0--3. 
  Bottom: Expert consensus filtering criteria to assure quality.}
  \label{fig:bbps_intro}
  \vspace{-0.1in}
\end{figure}

The Boston Bowel Preparation Scale (BBPS) has emerged as the most widely adopted metric for quantifying colon cleanliness, providing a standardized, segment‐by‐segment score from 0 (unprepared) to 3 (excellent cleansing) (see Figure~\ref{fig:bbps_intro}). 

However, manual BBPS assessment suffers from inter‐observer variability, is difficult to harmonize across institutions, and demands extensive clinician training—factors that consume valuable time and resources. Automating BBPS scoring not only promises to streamline the workflow and reduce resource expenditure but also helps eliminate subjective bias and ensure consistent evaluations across different hospitals, making it a pivotal goal for next‐generation computer‐aided diagnostic systems.

Despite significant interest in automated BBPS scoring, two primary challenges remain. First, there is a dearth of diverse, high‐quality, standardized datasets tailored for BBPS evaluation. To our knowledge, the only publicly available colonoscopy dataset (NERTHU~\cite{Pogorelov:2017:NBP:3083187.3083216}) contains images from just ten  subjects, a scale far below what real‐world variability demands. This lack of representative data hinders both effective model training and robust performance assessment. Second, medical imaging tasks require exceedingly high accuracy to satisfy clinical standards, yet assembling large, annotated endoscopic image corpora is inherently difficult. Training deep networks from scratch on limited samples typically yields suboptimal performance (~\cite{vabalas2019machine}). Prior work (e.g., Singh et al.~\cite{singh2023automated}) has applied conventional pretrained models, but their transferability from natural images to the endoscopic domain is limited, leading to degraded performance in bowel‐cleanliness assessment.

To address the first challenge, we have constructed a more diverse and rigorously labeled dataset of 2,240 colonoscopy images from 517 subjects, each annotated with BBPS scores. To ensure labeling consistency and mitigate subjectivity, three experienced endoscopists independently scored every image; only those with unanimous agreement were retained. This high‐quality dataset enables both effective model training and reliable evaluation.

To overcome the second challenge, we propose a region-aware feature fusion network, which takes advantage of the strong generalization capabilities of large‐scale pretrained foundation models. Specifically, we adopt the CLIP model~\cite{radford2021learning} and apply adapter‐based transfer learning to adapt it to the endoscopic domain, thus avoiding the need to train from scratch on limited medical data. Moreover, given that the BBPS score is highly correlated with the amount of stool observed in colonoscopy images, we introduce a dedicated fecal‐feature extraction branch: by utilizing anchors to extract fine-grained reign features and incorporating textual priors that describe stool characteristics, we avoid the overhead of retraining a separate segmentation network. The extracted fecal features are fused with visual embeddings to enhance BBPS prediction. Experiments on both the HDFD dataset and the NERTHU dataset demonstrate that our framework achieves state‐of‐the‐art performance in automated bowel‐preparation scoring. In summary, our contributions are threefold:
\begin{itemize}
    \item[1)] A high‐quality, diverse BBPS‐annotated colonoscopy image dataset(HDFD) is constructed comprising 2,240 images with expert‐validated labels.
    \item[2)] A CLIP‐based BBPS scoring architecture "Region-Aware Feature Fusion Network" is proposed, incorporating adapter‐based transfer learning~\cite{houlsby2019parameter} and a region-aware novel fecal‐feature extraction branch.
    \item[3)] Extensive experiments are conducted on multiple datasets to validate the effectiveness of the pretrained backbone, transfer adapters, and fecal‐feature module in improving automated BBPS accuracy. 
\end{itemize}

\vspace{-0.1in}
\section{Related Works}

\subsection{BBPS Scoring and Clinical Significance}

The Boston Bowel Preparation Scale (BBPS), first proposed by Lai et al.~\cite{lai2009boston}, has been widely adopted as a reliable metric for bowel cleanliness assessment, demonstrating robust inter- and intra-observer consistency ~\cite{calderwood2010comprehensive}. Clinical validation studies support BBPS scores $\ge2$ as sufficient for a high-quality colonoscopy~\cite{clark2016quantification} and use them as a basis for recommending 10-year follow-up intervals~\cite{calderwood2014boston}. Furthermore, BBPS exhibits strong cross-cultural reliability, with high agreement reported in Chinese clinical settings \cite{gao2013pilot}. Prior research also found significant associations between lower BBPS scores and increased polyp miss rates\cite{kluge2018inadequate}, emphasizing the scale’s diagnostic importance.

Several interventions have aimed to improve bowel preparation quality as measured by BBPS. For example, cartoon-based educational guides~\cite{tae2012impact} and inpatient information booklets~\cite{ergen2016providing} have both demonstrated statistically significant improvements in patient compliance and cleanliness outcomes.

Automated BBPS scoring systems have recently emerged to address subjectivity and variability in manual assessment. Hossain et al.\cite{hossain2023deeppoly} applied Vision Transformers (ViTs) to classify endoscopic images, while Sharma et al.\cite{sharma2022ensemble} proposed CNN ensemble models to address limited data issues. Lee et al.~\cite{lee2022artificial} introduced a two-stage CNN pipeline, achieving high agreement with expert annotations.
This paper addresses these challenges by leveraging pretrained vision-language models and region-aware feature fusion for automated BBPS scoring.

\vspace*{-2mm}
\subsection{Transfer Learning for Medical Image Analysis}

Given the scarcity and specialized nature of medical image data, transfer learning has become essential to mitigate overfitting. Prior studies demonstrated successful transfer learning in related medical tasks, such as VGG-based polyp classification~\cite{ribeiro2016exploring} and gland segmentation using minimal network architectures~\cite{graham2019minimal}.

Inspired by these findings, our method specifically employs a pre-trained ViT-B/16 backbone derived from PMC-CLIP~\cite{lin2023pmc}, limiting fine-tuning to classifier parameters for improved data efficiency. This approach aligns with broader trends in multimodal learning and vision-language pre-training (VLP), where models like VLMo~\cite{bao2022vlmo} leverage shared-attention transformer designs to achieve strong performance in low-data cross-modal settings.

More generally, deep multimodal frameworks~\cite{ramachandram2017deep} enable robust fusion of heterogeneous data modalities, improving generalization and interpretability—two essential features for trustworthy medical AI systems. Additionally, the field of no-reference medical image quality assessment (IQA), widely used in MRI and CT~\cite{chow2016review}, highlights domain-specific challenges that further support the need for interpretable and data-efficient designs such as ours.
This paper specifically focuses on exploring model transfer for automated BBPS scoring.

\section{HDFD Dataset Construction}
To build a more comprehensive evaluation benchmark, we constructed our own dataset,named HDFD. In this section, we detail the dataset creation process, the evaluation metrics we use, and a comparison with the public NERTHU dataset, demonstrating the necessity of our data and its advantages over existing public collections.

\vspace*{-2mm}
\subsection{Construction Process}
We adopted a retrospective approach, first gathering colonoscopy videos of 517 subjects from published sources. To prevent the dataset from containing overly similar frames, we sampled images at 0.5-second intervals and then semi-automatically removed any blurred frames. To minimize subjectivity in BBPS scoring and ensure high-quality annotations, three experienced colonoscopists independently rated each image with an integer score in {0, 1, 2, 3}, and only images with unanimous agreement were retained. This process yielded a dataset of 2,240 images, each labeled with a BBPS score.
\vspace{-0.1in}
\subsection{Evaluation \& Comparison}
Several key statistics were computed for our dataset and compared against those of NERTHU, the only publicly available benchmark, to demonstrate the necessity of constructing a new dataset and to highlight its advantages. Specifically, we evaluated the number of \textbf{subjects} from whom videos were collected, the overall dataset \textbf{scale} (total number of samples), \textbf{per-class sample counts}, image \textbf{resolution}, \textbf{intra-class distance}, and \textbf{inter-class distance}; the detailed results for both datasets are summarized in Table \ref{stat}.\par

Table \ref{stat} shows that our HDFD dataset expanding the source population from 10 to 517 patients markedly increases inter‐class distance (20.20 vs. 18.61), reflecting substantially richer sample diversity. Despite a comparable overall size (2 240 vs. 1 956 images) and balanced per‐class counts (560 vs. 500), this subject‐driven heterogeneity significantly enhances class separability and real‐world applicability.

\begin{table}[ht]
  \centering
  \footnotesize
  \label{dataset0}
  \begin{tabular}{ccccccc}
    \toprule
    Dataset & Subject & Scale & per-class samples & resolution& Intra-Dist. & Inter-Dist.  \\ \hline
    NERTHU & 10 & 1956 & 500 & 720*576 &15.24&18.61 \\
    HDFD  (Ours)&  \textbf{517} &  \textbf{2240} &  \textbf{560} &  \textbf{1243*1052} & \textbf{18.81}& \textbf{20.20}\\
    \bottomrule
  \end{tabular}
\caption{Statistical comparison between NERTHU and the proposed HDFD dataset. Dist stands for Distance.}
\label{stat}
\vspace{-0.1in}
\end{table}

\begin{figure}[ht]
  \centering
  \includegraphics[width=0.80\linewidth]{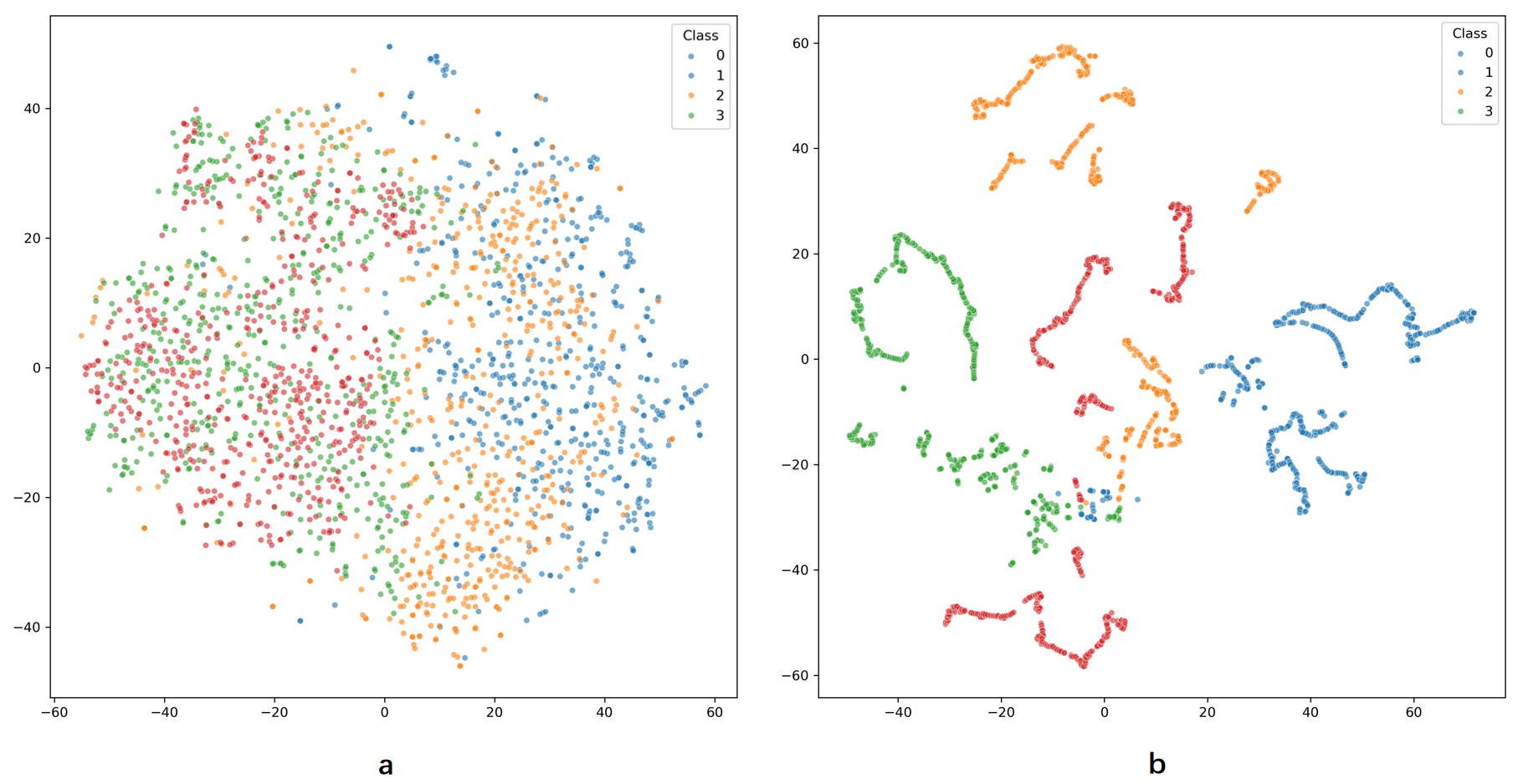}
  \caption{t-SNE visualization of feature distributions. Different colors indicate different BBPS scores. (a) Our HDFD dataset; (b) Public NERTHU dataset. }
  \label{fig:tsne_compare}
  \vspace{-0.1in}
\end{figure}

To more intuitively demonstrate that our dataset exhibits greater diversity and thus more accurately reflects real-world conditions, we extracted features from all samples using pretrained ResNet\cite{he2016deep} and visualized them with t-SNE, as shown in Fig.~\ref{fig:tsne_compare}. In the figure, (a) shows the distribution for our HDFD dataset, while (b) depicts the external NERTHU dataset.

From the figure, it is observed that: In the NERTHU visualization, the four BBPS scores form well-separated clusters with sharp boundaries and sparse inter-cluster distributions, reflecting excellent discriminability but also suggesting an idealized sample composition lacking the gradual ambiguity found in real clinical scenarios. In contrast, our dataset exhibits a more continuous sample distribution with noticeable overlap—particularly between scores 1 and 2—indicating smoother transitions at class boundaries. This blurring of category borders more closely mirrors the subjective judgment process of endoscopists and the inherent uncertainty in real-world BBPS scoring. In other words, our dataset better captures the nuanced, fuzzy boundaries of clinical practice, which should help models generalize under complex decision surfaces.\par

\vspace*{-2mm} 

\section{Framework}

\vspace*{-1mm} 
\subsection{Preliminaries}
We frame automated BBPS scoring as a four-class image-classification task.  In particular, each colonoscopy image $x\in\mathcal{X}\subset\mathbb{R}^{3\times H\times W}$ (of height $H$ and width $W$) must be assigned one of the four BBPS scores $y\in\mathcal{Y}=\{0,1,2,3\}$.  Let the full dataset be
$\;D=\{(x_i,y_i)\}_{i=1}^N$,
which we split into disjoint subsets $D_{\mathrm{train}},D_{\mathrm{val}}$ and $D_{\mathrm{test}}$ for model development, hyperparameter tuning, and final evaluation, respectively.  Our goal is thus to learn a network $f\colon\mathcal{X}\to\mathcal{Y}$ that—given a new image—correctly predicts its BBPS score.




\subsection{Method Overview}
\begin{figure}[ht]
  \centering
  \includegraphics[width=0.9\linewidth]{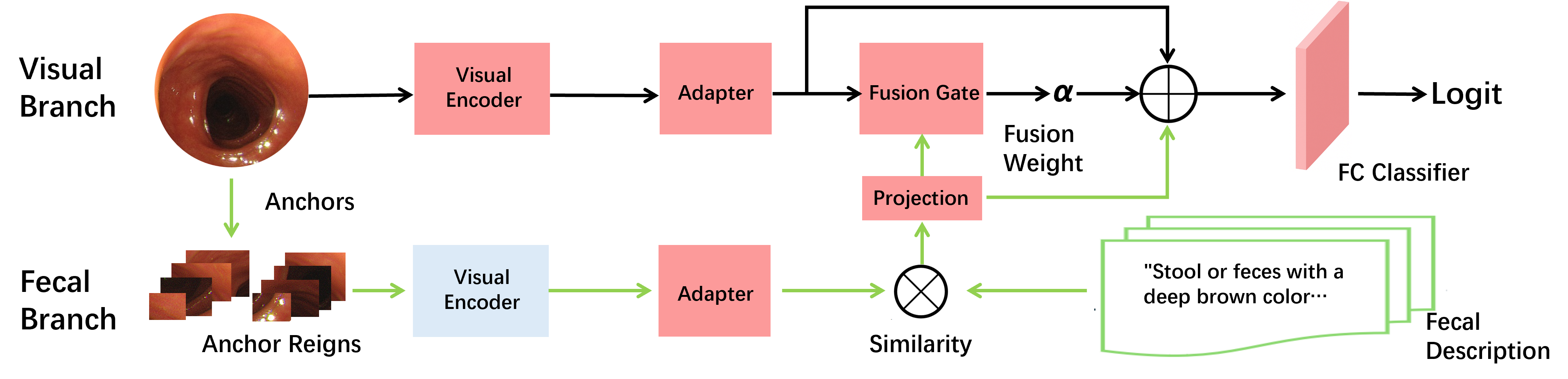}
  \caption{The architecture of proposed region-aware feature fusion framework. Red denotes trainable network components, while blue represents frozen parameters.}
  \label{fig:framework}

\end{figure}

Our framework, as illustrated in Figure~\ref{fig:framework}, adopts a dual-branch architecture consisting of a main branch and a stool-feature branch. The main branch combines a CLIP visual encoder with lightweight adapters and processes the entire image to extract global representations. In parallel, the stool-feature branch takes anchors of varying shapes as input and—guided by textual descriptions of stool—extracts localized fecal cues. Finally, we fuse these two feature streams using a gated-attention mechanism and feed the resulting representation into a classifier to produce the final BBPS score prediction. In the following, we first present the detailed design of our framework, and then describe the training and inference procedures.



\subsection{Framework}
\noindent\textbf{Visual Branch: Global Representation via CLIP and Adapter. } 
The input to the visual branch is the entire colonoscopy image. The goal of this branch is to extract a global representation that captures the overall cleanliness of the bowel. To achieve this, we utilize the CLIP's visual encoder as our backbone, defined as $f_{vit}(\cdot)$. However, considering our dataset is relatively small, directly fine-tuning large vision transformers is prone to overfitting. Therefore, to enhance the transferability of the model while preserving generalization from large-scale pretraining, we introduce an Adapter module after the CLIP encoder, defined as $g_{ada}(\cdot)$. The image is first passed through the ViT encoder to extract deep visual features and then refined through the adapter. The output $z_v$ is a global feature vector representing the entire image, which serves as the main input to the classification module. The overall formula for the visual branch is as follows:
\begin{equation}
z_v = g_{ada}(f_{vit}(x_i)), 
\label{eqa_v}
\end{equation}
where $x_i$ is the input image, $z_v$ is the output of visual branch.\par
\noindent\textbf{Fecal Branch: Localized Stool Feature Extraction via Anchors and CLIP.} To complement global features with localized details relevant to bowel cleanliness—especially residual stool—we introduce a fecal feature extraction branch. The input to this branch consists of multiple anchor regions sampled from the input image. These anchors are designed with various shapes and aspect ratios (e.g., 1:1, 2:1, 1:2, 3:1, 1:3) to capture diverse spatial patterns. The aim is to finely detect feces in small or elongated regions that may not be visible in the global view.
Each anchor region is first cropped and resized, then passed through a dedicated CLIP visual encoder and adapter—architecturally identical to those in the main visual branch but with independent parameters, denoted $f'_{\mathrm{vit}}(\cdot)$ and $g'_{\mathrm{ada}}(\cdot)$. Textual prompts (e.g., “yellow stool,” “residual feces”) are encoded via the CLIP text encoder. For each anchor, its visual embedding is compared against every prompt embedding $p$ to yield similarity scores that reflect the likelihood of stool presence. These scores are then aggregated into a single fecal feature vector $z_f$, which encapsulates all local stool-related cues. The overall formula for the feacal branch is as follows:
\begin{equation}
z_f = Sim(g'_{ada}(f'_{vit}(x_a)), p), 
\label{eqa_f}
\end{equation}
where, $x_a$ stands for the anchor region of the input image,  $p$ stands for the prompts features, $Sim$ stands for the similarity operation.\par
\noindent\textbf{Fusion Branch: Gated Fusion with Residual Emphasis.} 
To harmonize global visual context with localized fecal cues, gated fusion mechanism is utilized. Firstly, the fecal feature vector is linearly projected into the same embedding space as the global visual feature. Then, these two features are concatenated and passed through a lightweight gating network that produces a fusion weight vector via a sigmoid activation. The fused output feature $z_f$ is obtained by an element-wise weighted sum of the visual and projected fecal features using the fusion weights, preserving the primary global context while dynamically integrating only the most relevant local stool cues. The overall formula for fusion branch is as follows:
\begin{equation}
\begin{aligned}
\alpha &= f_{fusion}(z_v;FC(z_f)),\\
z_{all} &= \alpha \odot z_v + (1 - \alpha) \odot z_f,
\end{aligned}
\label{eqa_fusion}
\end{equation}
where $z_all$ is the fused output, $FC$ is the linear projection, $\alpha$ is the fusion weight.\par
Finally, the fused feature vector is passed through a fully connected classifier to produce classification logits.

\subsection{Training \& Inference}
\textbf{Training Procedure:} An input image $x_i$ is first processed by the global visual branch to produce the feature vector $z_v$ (Eq. \ref{eqa_v}). In parallel, anchor regions are cropped from $I$ and passed through the fecal branch to yield the feature vector $z_f$ (Eq. \ref{eqa_f}). These two streams are then fused according to Eq.\ref{eqa_fusion}, resulting in the combined representation $z$. The fused feature $z$ is fed into a fully connected classifier, which outputs logits for BBPS classification. A cross-entropy loss is computed between the logits and the ground-truth labels, and gradients are backpropagated to update all trainable parameters. To conserve memory, the CLIP visual encoder in the fecal branch remains frozen throughout training.

\noindent\textbf{Inference Procedure:}  
During inference, an input image is processed through the global and fecal branches (Eq. \ref{eqa_v} and Eq. \ref{eqa_f}), and fused via Eq. \ref{eqa_fusion} to produce the feature vector $z_all$. The classifier then computes logits $\boldsymbol{logit} = \mathrm{Classifier}(z_{all})$, and the predicted BBPS score is given:

\begin{equation}
\hat{y} = \arg\max_c \;logit,   
\end{equation}
All network parameters remain frozen during the entire validation and testing phase.

\par
\vspace*{-1mm}
\vspace*{-2mm} 
\section{Experiments}
\subsection{Experimental Setup}

\textbf{Implementation Details:} 
We train our model on a single NVIDIA GeForce RTX 4090 GPU. For the visual encoder, we adopt the ViT-B/16 backbone from CLIP, pre-trained on the DFN2B dataset. Within the adapter module, the intermediate layer’s hidden dimension is set to one-quarter of the input dimension, and we employ 180 anchors. The gating network is configured such that the output dimension of its first layer and the input dimension of its final layer both match the adapter’s output feature size.
Data augmentation follows the CLIP standard: random horizontal flips, random rotations, color jittering, and Gaussian blur. We optimize all parameters with AdamW, using a learning rate of $1e^{-5}$ for the CLIP visual encoder and $1e^{-3}$ for all newly introduced layers, with a weight decay of 0.01. Mixed-precision training is applied to accelerate convergence.
After each epoch, we evaluate on the validation set and retain the checkpoint achieving the highest validation performance for final testing.\par
\noindent\textbf{Baselines and Competitors:}  
To ensure a fair and consistent comparison, all experiments in this study are conducted under the same network architecture, using ViT-B/16 as the backbone. During training, the backbone is frozen, and only the classifier layers are updated. We evaluate multiple pre-trained models under this setting, including PMC-CLIP, OpenAI CLIP, and DFN2B.  In addition, to simulate a heterogeneous-source setting, we also evaluate OpenAI CLIP with an RN50 backbone under identical conditions. All pre-trained models are sourced from the \texttt{open\_clip}~\cite{ilharco_gabriel_2021_5143773} library or OpenAI, and the classifier is fine-tuned with an AdamW optimizer. \par
\subsection{Main Results}
To evaluate the effect of different pretrained vision-language models and their generalization ability in BBPS scoring, we compare multiple backbone architectures on two datasets: our proposed \textbf{HDFD dataset} and the publicly available \textbf{NERTHU dataset}.


\begin{table}[H]
  \centering
  \footnotesize
  \begin{tabular}{l c c c c c}
    \toprule
    Backbone (Model) & BBPS0 & BBPS1 & BBPS2 & BBPS3 & AVG.\ (\%)\\
    \midrule
    ViT-B/16 (PMC-CLIP) & 32.00 & 75.00 & 67.00 & 64.00 & 59.50 \\
    ViT-B/16 (OpenAI) & 73.00 & 91.00 & \textbf{95.00} & 92.00 & 87.75 \\
    ViT-B/16 (DFN2B) & 84.00 & 86.00 & 90.00 & 88.00 & 87.00 \\
    RN50 (OpenAI) & 73.00 & 56.00 & 48.00 & 88.00 & 66.25 \\
    ViT-B/32 (OpenAI) & 77.00 & 70.00 & 70.00 & 87.00 & 76.00 \\
    \textbf{Ours} & \textbf{86.00} & \textbf{91.00} & 91.00 & \textbf{96.00} & \textbf{91.00} \\
    \bottomrule
  \end{tabular}
  \caption{
 BBPS classification accuracy (\%) on the proposed HDFD dataset.
  }
  \label{tab:ours_internal_results}
  \vspace{-0.1in}
\end{table}
\begin{table}[H]
  \centering
  \footnotesize
  \begin{tabular}{l c c c c c}
    \toprule
    Backbone (Model) & BBPS0 & BBPS1 & BBPS2 & BBPS3 & AVG.\ (\%)\\
    \midrule
    ViT-B/16 (PMC-CLIP) & 100.00 & 100.00 & 100.00 & 0.00 & 65.17 \\
    ViT-B/16 (OpenAI) & 100.00 & 100.00 & 100.00 & 100.00 & 100.00 \\
    ViT-B/16 (DFN2B) & 100.00 & 100.00 & 100.00 & 100.00 & 100.00 \\
    RN50 (OpenAI) & 100.00 & 100.00 & 100.00 & 29.21 & 82.30 \\
    ViT-B/32 (OpenAI) & 100.00 & 100.00 & 100.00 & 58.43 & 89.61 \\
    \textbf{Ours} & \textbf{100.00} & \textbf{100.00} & \textbf{100.00} &\textbf{100.00} & \textbf{100.00} \\
    \bottomrule
  \end{tabular}
  \caption{
  BBPS classification accuracy (\%) on the public NERTHU dataset.
  }
  \label{tab:external_results}
\end{table}

From the results on two dataset it can be concluded that: On both dataset, the proposed method achieves the best overall accuracy of 91.00\% and 100\%, outperforming all baseline models, including DFN2B (87.00\%) and OpenAI CLIP ViT-B/16 (87.75\%). 1) On the HDFD dataset, our model achieves the highest scores in BBPS0 (86.00\%), BBPS1 (91.00\%), and BBPS3 (96.00\%), specifically outperforming the second-best models in BBPS0 and BBPS3 by +2.00\% (vs. DFN2B’s 84.00\%) and +4.00\% (vs.ViT-B/16’s 92.00\%), respectively. Since these two categories represent extreme bowel conditions (heavily contaminated and fully clean), their strong performance illustrates the benefit of incorporating region-aware fecal features. 2) For  BBPS2 (91.00\%), our method ranks second, only slightly behind the top scores (ViT-B/16's 95.00\% for BBPS2). This suggests that while fecal features contribute, BBPS2 classification may also rely on subtler visual cues such as mucosal visibility, where residual stool is present but does not obscure the lumen \cite{lai2009boston}. 3) On the external dataset, our method achieves 100.00\% overall accuracy, outperforming or matching all other models. While many baselines also reach 100.00\% on BBPS0–2, some still exhibit sharp drops on BBPS3 (e.g., RN50 drops to 29.21\%), suggesting Dataset2 has low intra-class diversity and overly separable categories. 4) This contrast highlights the importance of our dataset, where per-class accuracies vary more realistically and challenge models to learn finer-grained features. By offering greater sample diversity and closer alignment to clinical ambiguity, our dataset provides a more reliable benchmark for robust BBPS scoring.
\subsection{Ablation Studies}  
\textbf{Ablation of Modules:} 
To evaluate the impact of each module in our framework, we conduct an ablation study on the HDFD dataset. Firstly, the CLIP-base model uses the ViT-B/16 backbone with only the classification head fine-tuned. Secondly, Adapter module is added to enhance transferability, forming the Trans-base configuration. Finally, our full model further incorporates a fecal-feature branch to capture region-aware cues critical for accurate BBPS scoring.\par
From the Table \ref{tab:ablation}, it is observed that: 1) The proposed two key components are effective, changing the clip-base for Trans-base increases the average accuracy by 1.25\%, and our full method adds another 2.75\%. 2) Our approach not only achieves the highest mean accuracy but also delivers more stable results across BBPS 1–3 all exceed 90\%—showing that adding fecal features makes the classification more consistent.\par
\begin{table}[ht!]
  \centering
  \footnotesize
  \begin{tabular}{l c c c c c c c c}
    \toprule
    Method & FT & Adapter & Fecal-feature & BBPS0 & BBPS1 & BBPS2 & BBPS3 & AVG \\
    \midrule
    CLIP-base & 	\checkmark & - & - & 84.00 & 86.00 & 90.00 & 88.00 & 87.00 \\
    Trans-base & 	\checkmark  & 	\checkmark  & - & \textbf{91.00} & 78.00 &\textbf{92.00} & 92.00 & 88.25 \\
    Ours & 	\checkmark & 	\checkmark & 	\checkmark & 86.00 &\textbf{91.00}& 91.00 & \textbf{96.00} & \textbf{91.00} \\
    \bottomrule
  \end{tabular}
  \caption{
  Ablation study (\%) to verify the effectiveness of each component on the HDFD dataset. Adapter and Fecal-feature branches were selectively ablated. FT = fine-tuned. 
  }
  \label{tab:ablation}
\end{table}

\noindent\textbf{Number of anchors:} We also evaluated how the number of region proposals (anchors) in the fecal-feature extraction branch affects BBPS classification accuracy. In this experiment anchors were generated across multiple scales and aspect ratios (e.g., 8×8, 4×4 grids, and 2:1, 1:2) to capture diverse local patterns. Instead of uniformly adjusting anchor counts, we adopted a range-aware strategy: reducing sparse large-scale anchors or expanding high-performance regions observed in preliminary trials. This reflects practical trade-offs: very small anchors may be noisy, while overly large ones may lack specificity. Experimental results are shown in the Table \ref{tab:anchor_analysis}.\par

As shown in Table~\ref{tab:anchor_analysis}, fewer anchors result in incomplete spatial coverage and reduced accuracy. More anchors improve fine-grained perception, though excessive numbers may cause redundancy. We choose 180 anchors as the default setting, balancing accuracy (91.00\%) and efficiency.
\begin{table}[H]
  \centering
  \footnotesize
  \begin{tabular}{lccccccc}
    \toprule
    Anchors & 22 & 37 & 52 & 85 & 180 & 353 & 564 \\
    \midrule
    Accuracy (\%) & 87.25 & 86.75 & 86.25 & 87.00 & \textbf{91.00} & 89.00 & 89.25 \\
    \bottomrule
  \end{tabular}
  \caption{Classification Accuracy Under Different Anchor Numbers (HDFD Dataset).}
  \label{tab:anchor_analysis}
\end{table}
\section{Conclusion}

In this work, we proposed a CLIP-based region aware fecal feature fusion framework for automated BBPS scoring, which integrates adapter-based transfer learning and a novel fecal-feature extraction module. In addition, to support robust training and piratical evaluation, we constructed a high-quality BBPS-annotated dataset HDFD of 2240 images from 517 subjects. Extensive experiments on both public and the proposed HDFD datasets confirm that our method achieves state-of-the-art performance, demonstrating its effectiveness in addressing key challenges in bowel cleanliness assessment and advancing the automation of colonoscopy evaluation.
\newpage

\bibliography{egbib}
\end{document}